\documentclass[11pt,twoside]{article}
\usepackage{./asp2014}

\aspSuppressVolSlug
\resetcounters

\begin{document}

\title{On the formation of DA white dwarfs with low hydrogen contents: Preliminary Results.}
\author{Miller Bertolami, M. M.$^{1}$, Althaus, L. G.$^{1,2}$, and C\'orsico, A. H.$^{1,2}$
\affil{$^1$Instituto de Astrofísica La Plata, CONICET-UNLP, Paseo del Bosque s/n, (B1900FWA) La Plata, Argentina; \email{mmiller@fcaglp.unlp.edu.ar, althaus@fcaglp.unlp.edu.ar, acorsico@fcaglp.unlp.edu.ar}}
\affil{$^2$Facultad de Ciencias Astronómicas y Geofísicas, Universidad Nacional de La Plata, Paseo del Bosque s/n, (B1900FWA) La Plata, Argentina.}}

% This section is for ADS Processing.  There must be one line per author.
\paperauthor{Miller Bertolami, M. M.}{mmiller@fcaglp.unlp.edu.ar}{}{CONICET}{Instituto de Astrofísica de La Plata}{La Plata}{Buenos Aires}{B1900FWA}{Argentina}
\paperauthor{Althaus, L. G.}{althaus@fcaglp.unlp.edu.ar}{}{Universidad Nacional de La Plata-CONICET}{Instituto de Astrofísica de La Plata}{La Plata}{Buenos Aires}{B1900FWA}{Argentina}
\paperauthor{C\'orsico, A. H.}{acorsico@fcaglp.unlp.edu.ar}{}{Universidad Nacional de La Plata-CONICET}{Instituto de Astrofísica de La Plata}{La Plata}{Buenos Aires}{B1900FWA}{Argentina}

\begin{abstract}
Systematic photometric  and asteroseismological studies in the last
decade support the belief that white dwarfs in the solar
neighborhood harbor a broad range of hydrogen-layer contents. The
reasons behind this spread of hydrogen-layer masses are not understood
and usually misunderstood.
% This is particularly true in the case of
% winds. 
In this work we present, and review, the different mechanisms
that can (or cannot) lead to the formation of white dwarfs with a
broad range hydrogen contents.
\end{abstract}

\section{Introduction}

White dwarf stars (WD) are the most common end point of stellar
evolution \citep{2010A&ARv..18..471A}.
% Consequently, they play a key role in our quest for understanding 
% the structure and history of our Galaxy. 
In the most simple scenario, WDs are formed when the
progenitor stars lose their external envelopes at the end of the
Asymptotic Giant Branch (AGB) and contract at constant luminosity to
the WD cooling sequence. Within this simple picture stellar
evolution theory usually predicts the formation of WD with pure
hydrogen (H) atmospheres and with a total H content of about $M_{\rm
  H}^{\rm WD}\sim 10^{-3}$--$10^{-5} M_\odot$. About 80\% of the
spectroscopically identified WDs are characterized by H-rich
atmospheres (usually referred as DA WDs). Considerable progress in the
study of WDs has been possible because some of these stars pulsate.
The study of the pulsational pattern of variable DA WDs (DAVs) through
asteroseismology has provided valuable constraints to their mechanical
properties, such as the core composition, the outer layer chemical
stratification and the stellar mass. Early asteroseismological
fittings suggested a range of H layer masses between $10^{-4}$ and
$10^{-10}$ $M_\odot$ ---see \cite{2001ASPC..226..307B} and references
therein. The mass of the outer H layer is an important point regarding
the spectral evolution theory ---see \cite{2011ApJ...737...28B},
%\cite{2015ASPC..493...33B} 
and also Rolland et al. in these
proceedings). \cite{2008ApJ...672.1144T} presented a detailed
photometric study of the spectral evolution of cool WDs
($T_{\rm eff}<15000$K).  They determined the ratio of helium(He)- to
H-atmosphere WDs as a function of temperature and concluded
that $\sim 15$\% of the DA WD between 15000 and 10000 K will be
transformed into He-atmosphere, non-DA, WD at lower temperatures. From
a model of convective mixing these authors then concluded that 85\% of
DA WD harbor H-layer masses larger than $M_{\rm H}^{\rm WD}\sim
10^{-6} M_\odot$, with the remaining 15\% harboring less massive
H layers. The existence a range of H-layer masses was later confirmed
by the systematic studies of \cite{2009MNRAS.396.1709C} and
\cite{2012MNRAS.420.1462R} by asteroseismological means. In
particular, the study of \cite{2012MNRAS.420.1462R}, finds 77\% of
their DAV sample (34 out of 44) to harbor H-layer masses larger than
$M_{\rm H}^{\rm WD}= 10^{-6} M_\odot$, in good agreement with the
photometric determinations of \cite{2008ApJ...672.1144T}.

In this context, it is interesting to analyze the evolutionary
channels that can lead to low H contents in DA WDs. Already
\cite{1990ARA&A..28..139D} reviewed several possible channels for the
formation of DA WDs with low H contents. These channels were: a) The
action of winds during the stationary He burning phase after a late
thermal pulse \citep{1979A&A....79..108S, 1984ApJ...277..333I}; b) a
late thermal pulse during the WD cooling track, where the
H envelope is violently burned
\citep{1984ApJ...277..333I,1995LNP...443...48I}; and c) a
diffusion/self-induced nova in the WD cooling track
\citep{1986ApJ...301..164I}. The study of \cite{2011MNRAS.415.1396M}
showed that the diffusion-induced nova does not lead to a significant
decrement of the H content of WDs. On the contrary scenarios a) and
b), now called ``late'' and ``very late'' thermal pulses (LTP, and
VLTP respectively; \citealt{2001Ap&SS.275....1B}) have been proven to
significantly reduce the H content of WDs. The quantitative
exploration by \cite{2005A&A...440L...1A} and
\cite{2005BAAA...48..185M} showed that H contents as low as $M_{\rm
  H}^{\rm WD}\sim 10^{-7} M_\odot$ could be produced in LTPs. VLTPs, on the
other hand, have been usually understood to produce non-DA WDs \citep{1984ApJ...277..333I,
  2005A&A...435..631A,2006PASP..118..183W}.  Although numerical
simulations never predict the burning of the complete H content of the
star \citep{2006A&A...449..313M}, it has been argued that whatever
the traces of H that may be left by the VLTP they will very likely be shed
off by mass loss during the subsequent giant phase
---\cite{2006PASP..118..183W}. While this is a possibility it has to
be emphasized that, once the star is back on its giant phase, H is
diluted in the more massive convective envelope of the born again AGB
star ($\sim 10^{-3} M_\odot$). The star would need to lose all that
mass in order to get rid of its whole H content. In addition, the
study of \cite{2007MNRAS.380..763M} suggested that the amount of
H burned in low-mass VLTPs might be only a small fraction of the total
H content of the star. Consequently, DA WDs with low H contents could
be formed both through LTP and VLTP events.

 In this work we present preliminary explorations aimed at a
 quantitative assessment of the range of WD with low H contents predicted by
 different post-AGB channels in the light of state-of-the-art stellar
 evolution models \citep{2016A&A...588A..25M}.  Although it was
 already clear from early works  that the H content of the
 future WD cannot be arbitrarily reduced by winds, during the AGB or
 H-burning phases \citep{1971AcA....21..417P,
   1979A&A....79..108S, 1987ASSL..132..337S}, some misconceptions have arisen in recent works.
% \citep{2009MNRAS.396.1709C, 2016ApJ...821...27G}. 
Consequently, before presenting the predictions of the LTP and VLTP
channels we will discuss why enhanced mass loss on the AGB and/or in
H-burning post-AGB sequences cannot lead to DA WDs with low
H contents.

\section{The misconceptions: Winds on the AGB and H-burning post-AGB phases.}
%\subsection{Misconception 1: Differences in metallicity}
%\articlefigure[width=.75\textwidth]{standard.eps}{Fig:Met}{$M_H^{\rm
%    WD}-M_{\rm WD}$ relation of DA WDs resulting from H-burning
%  post-AGB sequences with different $Z_{\rm ZAMS}$
%  \citep{2016A&A...588A..25M}.}

\articlefigure[width=.75\textwidth]{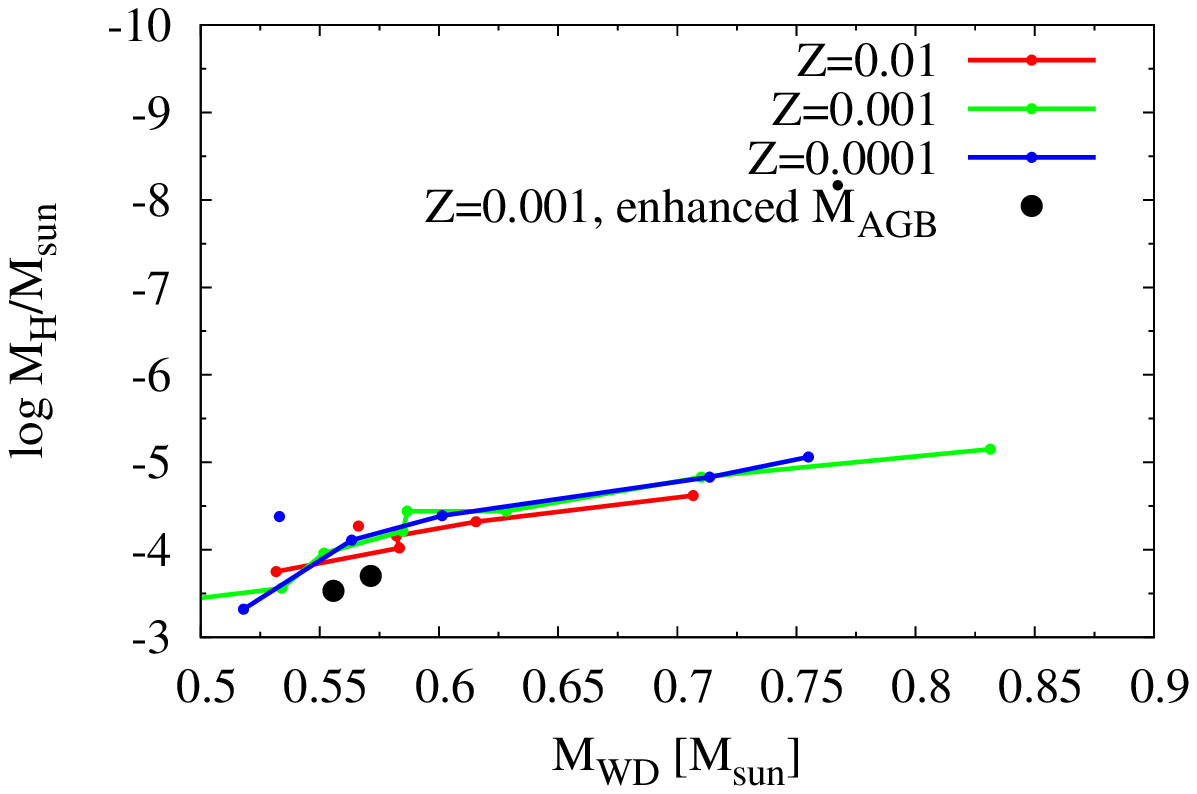}{Fig:AGB}{
  Solid Lines: $M_H^{\rm WD}-M_{\rm WD}$ relations of DA WDs resulting from
  H-burning post-AGB sequences with different $Z_{\rm ZAMS}$
  \citep{2016A&A...588A..25M}. Black dots:  $M_{\rm H}^{\rm WD}-M_{\rm WD}$ values of DA
  WDs constructed from an initially $M_{\rm ZAMS}=1.25 M_\odot$,
  $Z=0.001$ star by artificially increasing the AGB winds by a factor
  of 3 and 10.}

Fig. \ref{Fig:AGB} shows the dependence of the H content of the
resulting WD ($M_H^{\rm WD}$) with the initial progenitor metallicity
($Z_{\rm ZAMS}$) and final mass ($M_{\rm WD}$) as predicted for
H-burning post-AGB sequences by \cite{2016A&A...588A..25M}. As it is
apparent in Fig. \ref{Fig:AGB} the $M_H^{\rm WD}-M_{\rm WD}$ relation
of WD resulting from H-burning post-AGB sequences is not very
sensitive to the value of $Z_{\rm ZAMS}$ and restricted to the range
$10^{-3} M_\odot<M_{\rm H}^{\rm WD}< 10^{-5} M_\odot$.
% depending on the final mass $M_H^{\rm WD}$. 
Consequently, WDs evolved from H-burning
post-AGB objects show a rather tight $M_{\rm H}^{\rm WD}-M_{\rm WD}$
relation.

%In order to show the impact of winds on the AGB and H-burning post-AGB
%phases we have recomputed the sequence with $M_{\rm ZAMS}=1.25 M_\odot$,
%$Z=0.001$ presented in \citep{2016A&A...588A..25M} by artificially
%enhancing the winds both on the AGB and post-AGB.

%\subsection{Enhanced winds on H-burning post-AGB}

As noted by \cite{1971AcA....21..417P}, and also
\cite{1987ASSL..132..337S}, H-burning post-AGB models have a very
tight relationship between the effective temperature ($T_{\rm eff}$)
and the envelope mass  of the post-AGB object.
% This is similar to the better known tight $T_{\rm
%  eff}$-$M_{\rm env}$ relationship of objects on the horizontal branch
% \citep{2005slfh.book..149F} and it is valid for any structure with a
% dense core, a burning shell and a thin envelope in thermal equilibrium. 
%As a consequence,
As long as the envelope can be considered in thermal equilibrium, the
location of a H-burning post-AGB remnant on the HR-diagram is
independent of the mass loss history and only dependent on the value
of the envelope mass. As a consequence an enhancement in the post-AGB
winds does speed up the post-AGB evolution but does not reduce the
value of $M_H^{\rm WD}$. To check this argument we recomputed the
sequence with $M_{\rm ZAMS}=1.25 M_\odot$, $Z=0.001$ presented by
\cite{2016A&A...588A..25M} by artificially enhancing the winds on the
post-AGB phase.  Our computations show that even an enhancement of a
factor 100 in the post-AGB winds affects neither the $T_{\rm
  eff}$-$M_{\rm H}$ relation nor the final H content of the WD ($M_H^{\rm
  WD}$).

%\subsection{Enhanced winds on the AGB}

The intensity of the winds of AGB stars are not directly relevant
for the final H content of the WD. Fig. \ref{Fig:AGB} shows the impact
of increasing the AGB winds up to an order of magnitude. An
enhancement of AGB winds leads to a shortening of the AGB. With less
time for the H-free core to grow during the thermal pulses, enhancing
the winds on the AGB leads to a smaller final mass ($M_{\rm WD}$) for
the same initial mass ($M_{\rm ZAMS}$). Due to the tight $M_{\rm
  WD}$-$M_{\rm H}^{\rm WD}$ relation this implies that more intense winds on
the AGB lead to the formation of WD with higher H contents. This is
true also when looked at at the same value of the remnant mass ($M_{\rm
  WD}$). Models of the same final mass but shorter AGB lifetimes are
less compact and luminous and have, consequently, larger post-AGB
envelope masses \citep{1995A&A...299..755B, 2016A&A...588A..25M}.

Different AGB winds can indirectly affect the H content of the WD by
affecting the frequency of late thermal pulses. Depending on how AGB
winds are modulated with the surface temperature, brightness and
composition of the star, the occurrence of late thermal pulses can be
favored or not.  And, as shown in the next section, late thermal
pulses do indeed reduce the H content of the future WD.

\section{Late Flashers: A formation channel for DA WD with low H contents}
\articlefigure[width=.75\textwidth]{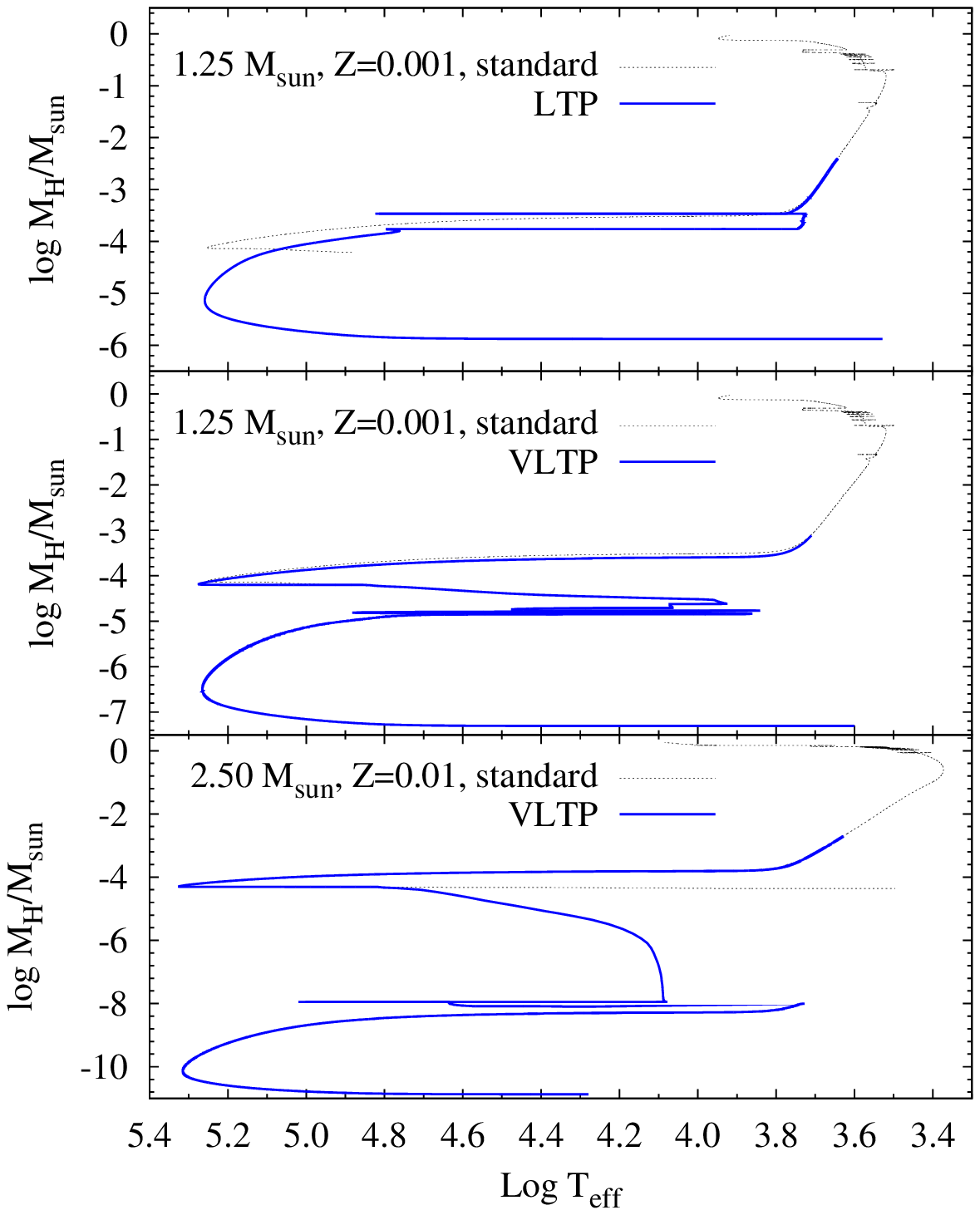}{Fig:VLTP-LTP}{Evolution of the H content
  ($M_{\rm H}$) of stars during LTP and VLTP events (dashed line) as
  compared with their canonical evolution in H-burning post-AGB
  sequences from the main sequence to the final WD value (dotted line).}

As discussed in the previous sections the occurrence of late thermal
pulses, or the action of winds during the He-burning phase after a
late thermal pulse can lead to an effective depletion of the H content
of the resulting WD stars. In order to quantitatively assess the final
H content expected from the VLTP and LTP events we have recomputed
some of the sequences presented in \cite{2016A&A...588A..25M}, by and
ad-hoc tuning of the mass loss during the departure from the AGB in
order to obtain a late thermal pulse. In particular we have computed
three LTP sequences corresponding to $(M_{\rm ZAMS},M_{\rm WD},Z_{\rm
  ZAMS})=(1.25 M_\odot,0.5900 M_\odot, 0.001)$, $(1.50 M_\odot,0.5961
M_\odot, 0.001)$, $(2.50 M_\odot,0.6200 M_\odot, 0.01) $ and two VLTP
sequences corresponding to $(M_{\rm ZAMS},M_{\rm WD},Z_{\rm
  ZAMS})=(1.25 M_\odot,0.5894 M_\odot, 0.001)$, $(2.50 M_\odot,0.6194
M_\odot, 0.01)$.  In Fig \ref{Fig:VLTP-LTP} we show the typical
evolution of the H content of post-AGB objects that undergo LTP and
VLTP events.

As discussed in \cite{2005A&A...440L...1A}, and shown in
Fig. \ref{Fig:VLTP-LTP}, after a LTP event the star returns to its
cool giant structure for a second time. If third dredge up
develops\footnote{The occurrence of third dredge up depends on the
  initial mass of the model and on the number of thermal pulses
  \citep{2016A&A...588A..25M}. } then H is diluted in the rather deep
convective envelope of the reborn red giant and the surface becomes
H deficient. Although no H is burnt in this phase, some H can be lost
due to stellar winds. Once the star departs from the AGB for a second
time, the envelope contracts and heats up. Then the H located in the
inner parts of the star is burnt and the total H content the present
simulations is reduced to $M_{\rm H}\sim 10^{-6}$--$ 10^{-7} M_\odot$
(see Fig.  \ref{Fig:Hcontent}). It is worth noting that final
H content of the WD could be higher if the LTPs would have taken place
closer to the AGB ---a scenario termed AGB final thermal pulse, see
\cite{2001Ap&SS.275....1B}.

%It is worth noting that final H-content of the WD after an LTP is
%given by the mass of the H-rich envelope at the moment of the
%He-flash. This is because after the LTP the H-rich envelope is diluted
%in the more massive (a few times $10^{-3} M_\odot$) H-free convective
%envelope of the reborn giant, leading to a H mass fraction in the
%envelope of $X_{\rm surf}^{post-LTP}$. As only the hydrogen
%located in the outermost $\sim 10^{-4.5}$ will survive once the star
%contracts to the a WD structure \citep{2005A&A...440L...1A}, the final
%H-content is $M_{\rm H}^{WD}\sim X_{\rm surf}^{post-LTP} \times
%10^{-4.5}$. Depending on the intensity of the dilution then LTP can lead to fin%al H-contents between $\sim 10^{-4.5}$ to $\sim 10^{-7}$

In the case of VLTP events, the H content of the models is also
reduced by the violent H flash that follows the thermal pulse ---see
\cite{2006A&A...449..313M} for a very detailed description of the
event. After the post-VLTP object returns to the AGB, the evolution
proceeds as in the LTP case. The remaining H is diluted in the
convective envelope of the reborn AGB star and, when the star departs
from the AGB for a second time H is burned again in the inner parts of
the envelope as it heats up. Interestingly, and in agreement with our
previous works \citep{2007MNRAS.380..763M} we find that the H flash
only burns part of the available H in low-mass VLTPs ($M_{\rm
  WD}\lesssim 0.6 M_\odot$). This is because in low-mass VLTP events
burning only a fraction of the available H injects enough energy in
the envelope of the star to expand it to its giant dimensions. On the
other hand, in more massive remnants undergoing VLTP events, burning
the whole H content is not enough to expand the envelope and H burning
only stops when the He-flash driven expansion finally pushes the star
back to the AGB.  As a consequence, in this preliminary explorations
we find in VLTP events that the final H content of the star strongly
depend on this mass threshold, being of the order of $M_{\rm H}\sim
10^{-11}M_\odot$ for the more massive sequence and much closer to that
expected in LTP events for the less massive ones ($M_{\rm H}\sim
10^{-7.3}M_\odot$, see Figs. \ref{Fig:VLTP-LTP} and \ref{Fig:Hcontent}).

\section{Discussion and future work}
\articlefigure[width=.75\textwidth]{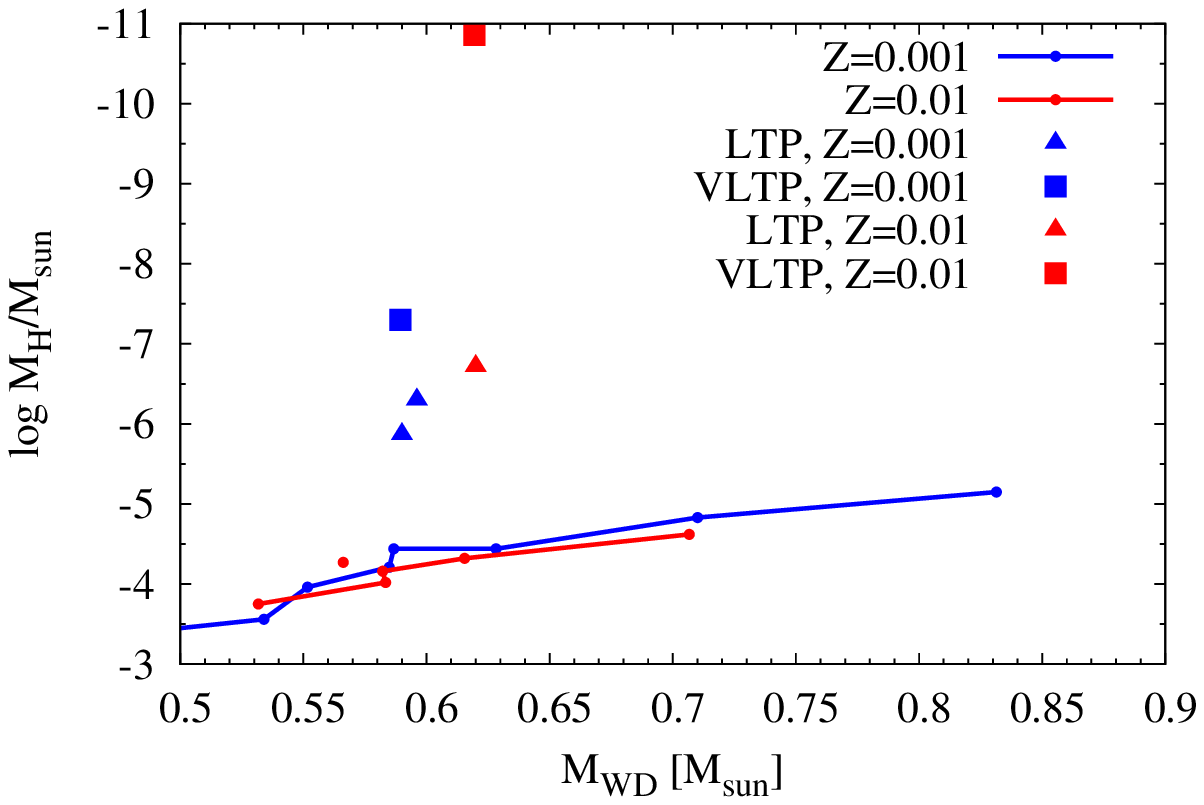}{Fig:Hcontent}{H content
  of the WDs predicted by VLTP and LTP channels as compared with the
  canonical predictions of H-burning post-AGB sequences for two
  different initial metallicities. }

In this preliminary study we have shown that both flavors of late
thermal pulses (VLTPs and LTPs) can lead to the formation of DA WDs
with low H contents (see Fig. \ref{Fig:Hcontent}). In addition we have
confirmed that uncertainties in AGB and post-AGB winds during the
H-burning phase are not directly relevant for the formation of DA WDs
with low H contents. In the case of LTPs our simulations predict
values as low as $M_{\rm H}^{\rm WD}\sim 10^{-7} M_\odot$. As the
amount of H remaining in this scenario is tightly connected to the the
mass of the H-rich envelope at the moment of the He-flash, we expect a
smooth transition from the canonical H-burning post-AGB value
(Fig. \ref{Fig:AGB}) down to $M_{\rm H}^{\rm WD}\sim 10^{-7}
M_\odot$. In addition, we have shown that current VLTP computations
predict a bi-modal distribution for the value of $M_{\rm H}^{\rm
  WD}$. With VLTP remnants of masses $M_{\rm WD}\gtrsim 0.6 M_\odot$
resulting in $M_{\rm H}^{\rm WD}\sim 10^{-11}M_\odot$ and less massive
VLTP remnants harboring H contents just below the LTP predictions
($M_{\rm H}\lesssim 10^{-7}M_\odot$). Future work will expand these
preliminary computations and explore the role of both remnant masses
and winds during the He-flash driven evolution.

%Explore the impact of winds, expand the grid...

\acknowledgements Part of this work was supported by ANPCyT through
grant PICT-2014-2708 from FonCyT and grant PIP 112-200801- 00940 from
CONICET. M3B thanks the Alexander von Humboldt Foundation for a Return
Fellowship. M3B and AHC thank the organizers for waiving their
respective registrations fees.

%\bibliography{editor}  % For BibTex
%\end{document}

% For non-BibTex:

\end{document}